\documentclass[english]{article}
\usepackage[T1]{fontenc}
\usepackage[utf8]{inputenc}
\usepackage{babel}
\usepackage{esint}
\usepackage[unicode=true,pdfusetitle,
 bookmarks=true,bookmarksnumbered=false,bookmarksopen=false,
 breaklinks=false,pdfborder={0 0 1},backref=false,colorlinks=false]
 {hyperref}
\usepackage{breakurl}
\begin{document}

\title{Local and global information and equations with left and right invertible
operators in the free Fock space}

\author{Jerzy Hanćkowiak, Poland, UE, e-mail: hanckowiak@wp.pl}

\maketitle
(former lecturer and research worker of Wroclaw and Zielona Gora Universities)

\date{November 2011}
\begin{abstract}
The subject of the work is short description of huge amount of n-point
information (n-pi) about the system. Methods for solving equations
that satisfy such information are considered. Possible interpretation
of left and right invertible operators appearing in these equations
is also proposed. For local information, the creation and annihilation
operators satisfying the Cuntz relations are introduced. It is also
introduced the vector describing the local vacuum which completes
the equation for n-pi with a global information. An important components
of the work are examples of local operator-valued functions. 
\end{abstract}
\tableofcontents{}

\section{Introduction}

Quotations:

``Recently there have been ambitious attempts to ground all of physics
in information; in other words, to treat the universe as a gigantic
informational or computational process (Frieden, 1998). An early project
of this type is Wheeler’s ``It from bit’ proposal (Barrow, Davies
\& Harper, 2003). We might call this 'level inversion' since information
is normally regarded as a higher-level concept than, say, particles.''
from P.Davies article - ``The physics of downward causation''  

The letters and words arranged with them, numbers, variables, and
various symbols used in science - are used to describe the environment
and ourselves. We are talking about variables when they may take different
values, but about the time-dependent functions when the values that
take variables are ordered, tracked over time. In the case of large
amounts of data concerning various aspects of the described system,
it is convenient to introduce the generating vector whose components
are the data. Moreover, often, the components are functions not only
depending on time, but from other variables and so infinite, often
uncountable amount of data, information about the system, one must
be able to describe in a compact way not only to control them, but
to be able to pass them or allow to them access of others.

The information contained in text such as - official letter or a novel
or a description of physical or mathematical theory - can be described
using the appropriate function $\varphi(t,\vec{x})$ where t - is
a given time, and $\overrightarrow{x}-$a point in space or place
in the book. In the latter case, the value of function $\varphi$
is the given letter of the alphabet $\mathit{\mathcal{A}}$, punctuation
or spacing between words. In the case of a physical system, such as
electromagnetic field, $\varphi$ is a multi-functions $\varphi_{i}(t,\vec{x});i=1,...,N$
describing the components of the field, which is further, to save
the recording, will be denoted by $\varphi(\tilde{x})$. 

Often, too detailed information $\varphi(\tilde{x})$ describing the
system (gas in the vessel, book, picture) is not available or indicated,
and then we are satisfied by its averaged or smoothed quantity $<\varphi(\tilde{x})>$:
In other words, we are satisfied by $\varphi(\tilde{x})\rightarrow<\varphi(\tilde{x})>$.
In fact, such process is carried out by our senses, instruments or
our theories, which are tailored to our manual and intellectual capabilities.
However, it is something in human nature that we would like to know
what is behind the world only accessible to our senses and instruments
and therefore so often what is really $<\varphi(\tilde{x})>$ is denoted
by $\varphi(\tilde{x})$ which need to be discovered. Here it is worth
mentioning the possibility of a completely crazy idea that $\tilde{x}$
may not have nothing to do with time and space, and appropriate averaging
only may have such relationship, see \cite{Heller (2008)}, page 413. 

It is not excluded that even at the classical level, there is a whole
hierarchy of descriptions in which each level is obtained by proper
filtration related to available instruments and theories. In other
words, the increasingly lower levels need not to be associated with
more subtle molecular structure of matter but rather with very sensitive,
ill posed description of the problem. Just as there is no point to
make ever finer triangulation of the measured surface, where individual
measurements are subject to a specific error, so it does not make
sense to look for increasingly accurate solutions to the Navier-Stocke
equations if we do not know the exact initial and boundary conditions. 

And now there is really a new thing that requires a new paragraph:
the value of $\varphi$ at point $\tilde{x}$, which could mean some
local feature of the system under test, such as the position of the
i-th item, and especially the change of $\varphi$, depends on other
system elements. And because this relationship is generally weakens
with distance, non-linearity appears here. The equations describing
the system are generally nonlinear. And what does this mean for the
smoothed or averaged quantities? It turns out that this leads to the
need to consider besides quantity $<\varphi(\tilde{x})>$ - averaged
products - $<\varphi(\tilde{x}_{1})...\varphi(\tilde{x}_{n})>;n=1,...,\infty$,
so to get a complete, infinite set of equations, which in previous
work we called - \textit{n-point (local) information (n-pi) of the
system}. 

The amount of local information about some physical systems or other
complex systems, such as economic, can not be written as a column
or row, used in the case of vectors with a finite or countable number
of components. Assuming that each of the n-pi is the coefficient standing
at the appropriate base vectors: $|\tilde{x}_{1},\cdots,\tilde{x}_{n}>\equiv|\tilde{x}_{(n)}>$
(Dirac notation), we use the symbols of the sum and integral to introduce
the vector | V> generating all these information:

\begin{equation}
|V>=<>|0>+\sum_{n=1}^{\infty}\int d\tilde{x}_{(n)}<\varphi(\tilde{x}_{1})...\varphi(\tilde{x}_{n})>|\tilde{x}_{(n)}>\label{eq:1}
\end{equation}
where coefficient standing at the vector |0> does not contain any
local information about the system. 

The use of one quantity, which is a vector |V>, instead of an infinite
number of n-pi, is a convenient simplification, and therefore the
operations referred to n-pi we also try to describe (express) with
a generating vector | V>. In this and previous works, we try to express
them by means of creation and annihilation operators which satisfy
the Cuntz relations,\ref{eq:5}. This leads to a very simple structure
for orthogonal basis vectors $|\tilde{x}_{(n)}>$, see Eq.\ref{eq:3},
which leads to a simple relations between the generating vector |
V> and n-pi:

\begin{equation}
<\tilde{x}_{(n)}|V>=<\varphi(\tilde{x}_{1})...\varphi(\tilde{x}_{n})>\label{eq:2}
\end{equation}

The use of generating vectors instead of generating functionals (fg),
see {[}Rzewuski 1969{]}, does not require assumptions about the convergence
of the respective sums. We will also assume that the n-pi $<\varphi(\tilde{x}_{1})...\varphi(\tilde{x}_{n})>$
are the usual functions, rather than generalized functions and, despite
use of the symbol of the integral, $\int$, the value of multicomponent
symbols $\tilde{x}$ are discrete, simply because to remember that
we do not exclude the continuous case. In other words, we assume some
regularization of considered theory. 

The set of vectors \ref{eq:1} constructed with the help operators
fulfilling the Cuntz relations (\ref{eq:5}) is called the \textit{free
(super, full) Fock space}. In this space the equation for the generating
vector |V> is constructed by means of one-side invertible operators
with explicitly constructed inverse operators. This allows us to construct
a suitable projection operators by which we express the general solution.
But a more important and difficult is a choice of physical solutions
which we try to obtain by additional assumptions expressed, for example,
by the perturbation or closure principles. It is worth noting that
the free Fock space allows us to describe the exact and averaged differential
equations, both in the commuting and noncommuting cases. 

Novelty of the submitted work, which can be read independently from
others author papers, is the in-depth understanding of the developing
formalism, in particular the role of one-side invertible operators
and the introduction of a new class of nonpolynomial operators describing
dynamics of the system, for which is given an appropriate perturbation
calculations. And although still not laid the appropriate equations
describing the dynamics of the formation of novels or other products
of humanism, the fact that the even novel events can be described
using the appropriate field $\varphi(\tilde{x})$ and that at the
perception of images, writing or reading novel in some sense - we
use filtration $<\varphi(\tilde{x})>$, leads us to believe that in
this ever-changing, fluid reality, (Zygmund Bauman), situation may
change. Notwithstanding this, the equations that we are considering
are already in operation in many fields of science and technology.

In a certain sense, by introducing the local and global entities,
we touched on the philosophical problem - the top-down causation,
see \cite{Craver (2006)}, getting a new look on spooky forces exerted
by wholes upon their components! In fact, the global variables are
integrated in some way fields. They can be trated as parameters of
theory. One can derive equations upon them and then it would means
that our ignorance about the system is not arbityrary but is limited
by solutions to these equations. To such conclusion we should come
if we want to keep in mind other useful computation assumptions like
the perturbation or closure principles. If in addition it would appear
that this is the only way of computation success, we would have to
assume that only certain ways of filtering leads to new information
about the system. In this sense we can say that our ignorance about
the system is quantized or rather limited!

\section{Operators that create and annihilate (local) information and vectors
that represent ``nothingness''}

In many cases the components of a vector have direct physical interpretation.
For example, the components of the radius vector can be interpreted
as projections of the position of the material point on the appropriate
base vectors. In quantum mechanics (QM) the physical interpretation
have got components as well as base vectors, which are eigenvectors
of appropriate Hermitian operators like the Hamilton operator or the
momentum operator. \textbf{To descsibe local information about the
system we do not need Hermitian operators}: That what we need are
one-side invertible operators. Among these operators are particularly
important cration and annihilation operators. With the help of these
operators and the ``vacuum'' vector |0> we create base vectors: 

\begin{equation}
|\tilde{x}_{(n)}>=\hat{\eta}^{\star}(\tilde{x}_{1})\cdots\hat{\eta}^{\star}(\tilde{x}_{n})|0>\label{eq:3}
\end{equation}
 Here, from definition, the operator $\hat{\eta}^{\star}(\tilde{x})$
\textit{creates a local information} about the system at the space-time
point $\tilde{x}$. The operator which annihilates this information
is denoted by $\hat{\eta}(\tilde{x})$. We will assume that

\begin{equation}
\hat{\eta}(\tilde{x})\hat{\eta}^{\star}(\tilde{x})=\hat{I}\label{eq:4}
\end{equation}
 where $\hat{I}$ is the unit operator in the space of generating
vectors. From the one side-invertible point of view, the property
(\ref{eq:4}) allows to call the operator $\hat{\eta}(\tilde{x})$
a right invertible operator, or, more rarely - a \textit{derivative},
and the operator $\hat{\eta}^{\star}(\tilde{x})$ - a left invertible,
or, a right invers to the operator $\hat{\eta}(\tilde{x})$. 

Additional assumption: 

\begin{equation}
\hat{\eta}(\tilde{x})\hat{\eta}^{\star}(\tilde{y})=\delta(\tilde{x}-\tilde{y})\cdot\hat{I}\label{eq:5}
\end{equation}
leads to equality:

\begin{equation}
\hat{\eta}(\tilde{x})|\tilde{x}_{1},\tilde{x}_{2},...,\tilde{x}_{n}>=\delta(\tilde{x}-\tilde{x}_{1})|\tilde{x}_{2},...,\tilde{x}_{n}>\label{eq:6}
\end{equation}
which, for the operator $\hat{\eta}(\tilde{x})$, justify the name
- the \textit{annihilation operator}. From (\ref{eq:6}) we also have:

\begin{equation}
\int d\tilde{x}\hat{\eta}(\tilde{x})\hat{\eta}^{\star}(\tilde{y})=\int d\tilde{y}\hat{\eta}(\tilde{x})\hat{\eta}^{\star}(\tilde{y})=\hat{I}\label{eq:7}
\end{equation}

Using base vectors (\ref{eq:3}) and (\ref{eq:5}), one can show that

\begin{equation}
\int d\tilde{x}\hat{\eta}^{\star}(\tilde{x})\hat{\eta}(\tilde{x})=\hat{I}\label{eq:8}
\end{equation}
Relations (\ref{eq:5}) in physical literature are called the \textit{Cuntz
relations}. To be in agreement with the usual restrictions imposed
upon the vectors |0> and <0|:

\begin{equation}
\hat{\eta}(\tilde{x})|0>=0,\;<0|\hat{\eta}^{\star}(\tilde{x})=0\label{eq:9}
\end{equation}
where <0| means a conjugate vector to the vector |0> on which operators
act from their left hand side, we modify Eq.\ref{eq:8} as follows:

\begin{equation}
\int d\tilde{x}\hat{\eta}^{\star}(\tilde{x})\hat{\eta}(\tilde{x})=\hat{I}-|0><0|\Leftrightarrow\hat{I}=\int d\tilde{x}\hat{\eta}^{\star}(\tilde{x})\hat{\eta}(\tilde{x})+|0><0|\label{eq:10}
\end{equation}
with restriction <0|0>=1. 

Eqs (\ref{eq:9}) can be interpreted in the following way: the operator
$\hat{\eta}(\tilde{x})$ acting on the vacuum vector |0>, which is
related only to global information about the system, see \cite{Hanckow 2011},
destroys this information completely and this fact is represented
by the zero vectors $\vec{0}\equiv0$ or $\overleftarrow{0}\equiv0$.
The same applies to the operators $\hat{\eta}^{\star}(\tilde{y})$
acting on the left. These two kind of vectors represent \textit{nothingness
}(no local information about the system). 

With Cuntz relations (\ref{eq:5}) and assumptions (\ref{eq:9}),
which means that the vectors \ref{eq:3} are orthonormal, it is easy
to prove:

\begin{equation}
<0|\hat{\eta}(\tilde{x}_{1})\cdots\hat{\eta}(\tilde{x}_{n})|V>=<\varphi(\tilde{x}_{1})...\varphi(\tilde{x}_{n})>\label{eq:11}
\end{equation}
and that

\begin{equation}
<0|V>=<>\label{eq:12}
\end{equation}

The following operators called projectors are also useful:

\begin{equation}
\hat{P}_{n}=\int d\tilde{x}_{(n)}\hat{\eta}^{\star}(\tilde{x}_{1})\cdots\hat{\eta}^{\star}(\tilde{x}_{n})|0><0|\hat{\eta}(\tilde{x}_{n})\cdots\hat{\eta}(\tilde{x}_{1});\;\hat{P}_{0}=|0><0|\label{eq:13}
\end{equation}

Now we have

\begin{equation}
|V>=<>|0>+\sum_{n=1}\hat{P}_{n}|V>\label{eq:14}
\end{equation}

\section{Equations for n-pi. An important modification}

We postulate the following linear equations for the n-pi $<\varphi(\tilde{x}_{1})...\varphi(\tilde{x}_{n})>$which
by means of the generating vector |V> can be described in the following
way:

\begin{equation}
\left(\hat{L}+\lambda\hat{N}+\hat{G}\right)|V>=\hat{P}_{0}\hat{L}|V>+\lambda\hat{P}_{0}\hat{N}|V>\equiv|0>_{info}\label{eq:15}
\end{equation}
where all operators $\hat{L},\hat{N},\hat{G}$ are linear operators
in the space of generating vectors (\ref{eq:1}) (only here we do
not use the name - Fock space - to emphasise that we do not assume
a norm space). Subsequent operators called here \textit{interaction
operators }can be related to subsequent terms in the equation:

\begin{equation}
L[\tilde{x};\varphi]+\lambda N[\tilde{x};\varphi]+G(\tilde{x})=0\label{eq:16}
\end{equation}
for the field $\varphi$, where $L$ depends in a linear way on the
field $\varphi$, and $N$ in a nonlinear way, see below. $G$ is
a given function. We call (\ref{eq:16}) the field equation. As an
example of such equation can be the Navier-Stock equations. There
is simple relation of functionals $L,N$ and function $G$ with operators
$\hat{L},\hat{N},\hat{G}$, see e.g. \cite{Hanckow 2011}. The most
general characteristic of these operators is that the operator 

\begin{equation}
\hat{L}=\int\hat{\eta}^{\star}(\tilde{x})L(\tilde{x},\tilde{y})\hat{\eta}(\tilde{y})d\tilde{x}d\tilde{y}+\hat{P}_{0}\label{eq:17}
\end{equation}
is a diagonal operator with respect to the projectors $\hat{P}_{n}$:

\begin{equation}
\hat{P}_{n}\hat{L}=\hat{L}\hat{P}_{n}\label{eq:18}
\end{equation}
where n=0,1,... Also, the operator 

\begin{equation}
\hat{N}=\int\hat{\eta}^{\star}(\tilde{x})N[\tilde{x};\hat{\eta}]d\tilde{x}+\hat{P}_{0}\hat{N}\label{eq:19}
\end{equation}
 is an upper triangular operator:

\begin{equation}
\hat{P}_{n}\hat{N}=\sum_{n<m}\hat{P}_{n}\hat{N}\hat{P}_{m}\label{eq:20}
\end{equation}
 where n=0,1,..., and the operator 

\begin{equation}
\hat{G}=\int\hat{\eta}^{\star}(\tilde{x})G(\tilde{x})d\tilde{x}\label{eq:21}
\end{equation}
 is a lower triangular operator:

\begin{equation}
\hat{P}_{n}\hat{G}=\hat{P}_{n}\hat{G}\hat{P}_{n-1}\label{eq:22}
\end{equation}
 for n=1,2,..., and, for n=0, $\hat{P}_{0}\hat{G}=0$. A more general
form of the lower triangular operator used in fact in quantum theories
has the following projection properties: 

\begin{equation}
\hat{P}_{n}\hat{G}=\sum_{m<n}\hat{P}_{n}\hat{G}\hat{P}_{m}\label{eq:23}
\end{equation}
 They express very deep, qualitative properties of described systems
like this that the system is immersed in a given external field, case
(\ref{eq:21}), or that the system is subjecting to certain constraints
that are implemented without the participation of reaction forces,
case (\cite{Hanckow (2011)};Sec.4.2).

\subsection*{An important modification}

As you can see from the above, Eq.\ref{eq:15} is identically satisfied
for its projection with the help of the projector $\hat{P}_{0}$.
This situation was caused by the fact that the operators $\hat{L}$
and $\hat{N}$ have been modified by adding to their expressions the
terms with projector $\hat{P}_{0}$, see (\ref{eq:17}) and (\ref{eq:19}).
In this way we obtained the operators, which in many cases are at
least right or left invertible, \cite{Hanckow (2010)}. Such a modification
of Eq.\ref{eq:15} does not affect the normal perturbation calculations
applied to Eq.\ref{eq:15}, but can influence other approach, see
\cite{Hanckow (2010)}, for which Eq.\ref{eq:15} without term $|0>_{info}$
is not complete. These are the mathematical reasons for the emergence
of vector $|0>_{info}$ in the right hand side of the Eq.\ref{eq:15}. 

Let us now look at this vector with a more physical point of view:
As it is known in many well-known equations of physics in their right
hand sides - the source of the fields that describe these equations
- appears. For example, see Maxwell's or Poisson's equations. Looking
at this spirit on the Eq.\ref{eq:15}, we can say that the vector
$|0>_{info}$ is a source of the vector |V>. In other worde, the source
of local information about the system is the vector containing global
information about the system, see also \cite{Hanckow (2010)-1} and
\cite{Hanckow (2010)}. 

It is also not excluded that such modification of equations on the
generating vector |V> can be used for rethinking some problems in
astrophysics and physics and perhaps to rescue certain useful paradigmas
like universality of law of physics, reductionist approach and so
on, \cite{Brooks (2009)}, so useful in previous development of science.

\section{Information overload and closure problem}

As we know many systems can in principle be described by means of
varies functions or fields. Even a novel or a picture can be described
in this way. If the system is too complex we have to use a certain
method of simplification of its description to get some practical
results. One such method is the filtering of information, which we
illustrate as follows:

\begin{equation}
\varphi(\tilde{x})\Rightarrow<\varphi(\tilde{x})>\label{eq:24}
\end{equation}

The process shown in (\ref{eq:24}) can be realized by omission, deletion
or averaging too detailed information. The difficulty that arises
here is that the functions $\varphi$, not as in the case of books
written or painted image, are not known. In many cases we know only
equations for these functions like Eqs (\ref{eq:16}) whose solution
is not a simple task and it is not even recommended due to consisting
too detailed information. In view of these difficulties in the nineteenth
century have emerged the idea to consider the equation for the 1-pi
$<\varphi(\tilde{x})>$. But here there is a new difficulty: the equation
for 1-pf usually contains other n-pfs, see Eq.\ref{eq:15} and Eq.\ref{eq:20},
and this difficulty is called the closure problem. 

The simplest recipe to solve the closure problem is to reject other
n-pi. But such method is justified only for very small value of the
coupling constant $\lambda$ if the generating vector |V> depends
analytically on $\lambda$. Here comes yet another difficulty, namely
the coupling constant $\lambda$, which describes the properties of
nonlinear interaction of the components of the system, is not generally
small. We have here such a situation that the nonlinearity problem
at the micro level is transformed into the closure problem at the
macro level. We have a paradoxical situation: what is simple, namely
the interaction between micro components of the system - leads to
a difficult closure problem. Since the nonlinearity in the micro-level
is common, because the decrease of interaction between components
of the system usually appear together with increase of their distances,
therefore the closure problem is common. Hence, different methods
of closing equations for n-pi are developed, see \cite{Hanckow (2010)},
\cite{Hanckow (2008)}, \cite{Hanckow (2007)} and literature there
cited.

\section{A few examples of operators $\hat{N}$}

Although n-pi $<\varphi(\tilde{x}_{1})...\varphi(\tilde{x}_{n})>$
are permutation symmetric, this does not mean that we must use the
symmetric base vectors $|\tilde{x}_{(n)}>$. Hence, instead of using
customarily accepted Heisenberg relations:

\begin{equation}
[\hat{\eta}(\tilde{x}),\hat{\eta}^{\star}(\tilde{y})]_{\mp}\propto\delta(\tilde{x}-\tilde{y})\hat{I}\label{eq:25}
\end{equation}
 which in result lead to permutation symmetric or anti symmetric bases
vectors $|\tilde{x}_{(n)}>$, we have used the Cuntz relations (\ref{eq:5}).
However, this has the advantage that the operators appearing in Eq.\ref{eq:15}
are right or left invertible. For example, 

\begin{equation}
\hat{N}=\int d\tilde{x}\hat{\eta}^{\star}(\tilde{x})\hat{\eta}^{2}(\tilde{x})+\hat{P}_{0}\int d\tilde{x}\hat{\eta}(\tilde{x})f(\tilde{x})\label{eq:26}
\end{equation}
appearing in the so called the Hurst model in Quantum Field Theory,
where $f$ is an arbitrary function, is the right invertible operator
with an easy constructed right inverse operator:

\begin{equation}
\hat{N}_{R}^{-1}=1/2\int d\tilde{y}\hat{\eta}^{\star}(\tilde{y})^{2}\hat{\eta}(\tilde{y})+1/2\int d\tilde{y}\hat{\eta}^{\star}(\tilde{y})\label{eq:27}
\end{equation}
 if $\int d\tilde{x}f(\tilde{x})=2$. Likewise, you can easily construct
a right inverse to the operator $\hat{L}$ and a left inverse to the
operator $\hat{G}$. It still does not solve the closure problem,
but can transform Eq.\ref{eq:15} into an equivalent manner and make
different regularization, see e.g., \cite{Hanckow (2010)}. Moreover,
the perturbation expansion with respect to the coupling constant $\lambda$is
gaining clarity, see previous author papers. 

At the end we give one more example of the operator $\hat{N}$ associated
with the non-linear part of Eq.\ref{eq:16}, which may prove useful
for further research in the proposed direction:

\begin{equation}
\hat{N}\equiv\hat{N}(\lambda_{2})=\int d\tilde{x}\hat{\eta}^{\star}(\tilde{x})\frac{H(\tilde{x})\hat{I}}{\hat{I}-\lambda_{2}\hat{\eta}(\tilde{x})}+\hat{N}\hat{P}_{0}\label{eq:28}
\end{equation}
 where $H(\tilde{x})$ is an arbitrary function and $\lambda_{2}$is
a new coupling constant. This operator generates other operators appearing
in Eq.\ref{eq:15}. For example, for $H=G$ and $\lambda_{2}=0$,
$\hat{N}=\hat{G}$. With expansion:

\[
\frac{\hat{I}}{\hat{I}-\lambda_{2}\hat{\eta}(\tilde{x})}=\hat{I}+\lambda_{2}\hat{\eta}(\tilde{x})+\left(\lambda_{2}\hat{\eta}(\tilde{x})\right)^{2}+\cdots
\]
we get the Hurst model supplemented with the $\varphi^{3}$- interaction
and other terms. Like in the polynomial case (Hurst model), it is
easy to construct a left inverse (sic) to the operator (\ref{eq:28}):

\begin{equation}
\hat{N}_{l}^{-1}\equiv\hat{N}_{l}^{-1}(\lambda_{2})=\int d\tilde{y}E(\tilde{y})\left(\hat{I}-\lambda_{2}\hat{\eta}(\tilde{y})\right)\hat{\eta}(\tilde{y})+\hat{P}_{0}\hat{N}_{l}^{-1}\label{eq:29}
\end{equation}
with restriction $\int d\tilde{x}E(\tilde{x})H(\tilde{x})=1$. 

The assumption: $\hat{P}_{0}\hat{N}=\hat{N}_{l}^{-1}\hat{P}_{0}=0$
does not contradict the equality:

\begin{equation}
\hat{N}_{l}^{-1}(\lambda_{2})\hat{N}(\lambda_{2})=\hat{I}\label{eq:30}
\end{equation}

\section{A generalization}

A generalization of the operator (\ref{eq:28}) is the formula 

\begin{equation}
\hat{N}\equiv\hat{N}(\lambda_{2})=\int d\lambda d\tilde{x}\hat{\eta}^{\star}(\tilde{x})\frac{h(\lambda)H(\tilde{x})\hat{I}}{\lambda\hat{I}-\lambda_{2}\hat{\eta}(\tilde{x})}+\hat{N}\hat{P}_{0}\label{eq:31}
\end{equation}
 leading to a more general expansions. A left inverse can be defined
by the expression:

\begin{equation}
\hat{N}_{l}^{-1}\equiv\hat{N}_{l}^{-1}(\lambda_{2})=\int d\mu d\tilde{y}E(\tilde{y})\frac{h^{-1}(\mu)\hat{I}}{\mu\hat{I}-\lambda_{2}\hat{\eta}(\tilde{y})}\hat{\eta}(\tilde{y})\label{eq:32}
\end{equation}
 where the function $h^{-1}$ is such defined that

\begin{equation}
\int d\mu\frac{h^{-1}(\mu)\hat{I}}{\mu\hat{I}-\lambda_{2}\hat{\eta}(\tilde{y})}=\left(\int d\lambda\frac{h(\lambda)H(\tilde{x})\hat{I}}{\lambda\hat{I}-\lambda_{2}\hat{\eta}(\tilde{x})}\right)^{-1}\label{eq:33}
\end{equation}
 At such a choice of the function $h^{-1}$ Eq.\ref{eq:30} is satisfied
and the operator $\hat{N}$ is a left invertible operator. 

Another example of an operator with easy constructed left inverse
is simply

\begin{equation}
\hat{N}(\lambda_{2})=\int d\tilde{x}\hat{\eta}^{\star}(\tilde{x})g\left(\lambda_{2}\hat{\eta}(\tilde{x})\right)H(\tilde{x})\label{eq:34'}
\end{equation}

In the paper \cite{Hanckow (2010)} we started from quite different
assumption, namely that the operator 

\begin{equation}
\frac{\hat{I}}{\lambda\hat{I}-\lambda_{2}\hat{\eta}(\tilde{x})}\equiv\left(\lambda\hat{I}-\lambda_{2}\hat{\eta}(\tilde{x})\right)^{-1}\label{eq:34}
\end{equation}
 is a right inverse to the operator $\left(\lambda\hat{I}-\lambda_{2}\hat{\eta}(\tilde{x})\right)$.
This means that operator $(\lambda\hat{I}-\lambda_{2}\hat{\eta}(\tilde{x}))$
is a right invertible operator. In this case, a right inverse operator
(\ref{eq:34}) can be chosen as a lower triangular operator and the
above property led to closing of the considered equations.

\section{A possible expansion of the generating vector |V>. The perturbation
principle}

Let us consider Eq.\ref{eq:15} with specified operators:

\begin{equation}
\left(\hat{L}+\lambda_{1}\hat{N}(\lambda_{2})+\hat{G}\right)|V>=\hat{P}_{0}|V>+\lambda_{1}\hat{P}_{0}\hat{N}(\lambda_{2})|V>\equiv|0>_{info}\label{eq:35}
\end{equation}
 where $\hat{L}$ is a right, and $\hat{N}$ is a left invertible
operator, given by Eq.\ref{eq:29} or Eq.\ref{eq:32}. Multiplying
Eq.\ref{eq:35} by a right inverse $\hat{L}_{R}^{-1}$ and using projector
$\hat{P}_{L}=\hat{I}-$$\hat{L}_{R}^{-1}\hat{L}$, which projects
on the null space of the operator $\hat{L}$, we can rewrite the above
equation in an equivalent way: 

\begin{equation}
\left(\hat{I}+\lambda_{1}\hat{L}_{R}^{-1}\hat{N}(\lambda_{2})+\hat{L}_{R}^{-1}\hat{G}\right)|V>=\hat{L}_{R}^{-1}|0>_{info}+\hat{P}_{L}|V>\label{eq:36}
\end{equation}

We will assume that solutions are symmetric:

\begin{equation}
|V>=\hat{S}|V>\label{eq:37}
\end{equation}
for example, the permutation symmetric. Because the operator $\hat{L}_{R}^{-1}\hat{G}$
is a lower triangular operator, the Eq.\ref{eq:36} can be equivalently
transformed further as follows:

\begin{eqnarray}
 & \left(\hat{I}+\lambda_{1}\left(\hat{I}+\hat{S}\hat{L}_{R}^{-1}\hat{G}\right)^{-1}\hat{S}\hat{L}_{R}^{-1}\hat{N}(\lambda_{2})\right)|V>=\nonumber \\
 & \left(\hat{I}+\hat{S}\hat{L}_{R}^{-1}\hat{G}\right)^{-1}\left(\hat{S}\hat{L}_{R}^{-1}|0>_{info}+\hat{S}\hat{P}_{L}|V>\right)\label{eq:38}
\end{eqnarray}
 where Eq.\ref{eq:37} was used. This equation can be a starting point
for the perturbation expansion with respect to the parameter $\lambda_{1}$
of the vector |V> satisfying the above equation. 

Now let us assume that the operator $\hat{N}$ allows the following
decomposition:

\[
\hat{N}(\lambda_{2})=\hat{N}(0)+\hat{N}_{1}(\lambda_{2})
\]
 In previous works we assumed that $\hat{N}$ is a right-invertible
operator and it was justified for a polynomial type of operators.
Now, we assume and at least formally justify that $\hat{N}$ is the
left invertible operator with a left inverse operator $\hat{N}_{l}^{-1}(\lambda_{2})$
, see Secs 5 and 6. From Eq.\ref{eq:35}, we get:

\begin{equation}
\left(\hat{N}_{l}^{-1}(\lambda_{2})(\hat{L}+\hat{G})+\lambda_{1}\hat{I}\right)|V>=\hat{N}_{l}^{-1}(\lambda_{2})|0>_{info}=0\label{eq:39}
\end{equation}
 We assume that

\begin{equation}
\hat{N}_{l}^{-1}(\lambda_{2})=\hat{N}_{l}^{-1}(0)+\hat{A}(\lambda_{2}),\;\hat{A}(\lambda_{2})\Rightarrow0,\; for\;\lambda_{2}\rightarrow0\label{eq:40}
\end{equation}
 The latter property makes that the terms containing the operator
$\hat{A}(\lambda_{2})$, for a small value of the coupling constant
$\lambda_{2}$, will be regarded as a perturbation. Substituting (\ref{eq:40})
into Eq.\ref{eq:39}, we get:

\begin{eqnarray}
 & \left\{ \left(\hat{N}_{l}^{-1}(0)+\hat{A}(\lambda_{2})\right)\hat{L}+\left(\hat{N}_{l}^{-1}(0)+\hat{A}(\lambda_{2})\right)\hat{G}+\lambda_{1}\hat{I}\right\} |V>=\nonumber \\
 & \left(\hat{N}_{l}^{-1}(0)+\hat{A}(\lambda_{2})\right)|0>_{info}\label{eq:41}
\end{eqnarray}
To use perturbation calculation, we have to transform the above equation
further to get an equation similar to Eq.\ref{eq:38}. For this purpose,
we multiply this equation by the operator $\hat{L}_{R}^{-1}\hat{N}(0)$.
We finally get:

\begin{eqnarray}
 & \left\{ \hat{I}+\hat{L}_{R}^{-1}\hat{N}(0)\hat{A}(\lambda_{2})\hat{L}+\hat{L}_{R}^{-1}\hat{N}(0)\left(\hat{N}_{l}^{-1}(0)+\hat{A}(\lambda_{2})\right)\hat{G}+\lambda_{1}\hat{L}_{R}^{-1}\hat{N}(0)\right\} |V>\nonumber \\
 & =\hat{\Pi}_{L}|V>+\hat{L}_{R}^{-1}\hat{N}(0)\left(\hat{N}_{l}^{-1}(0)+\hat{A}(\lambda_{2})\right)|0>_{info}\label{eq:42}
\end{eqnarray}
where the projector $\hat{\Pi}_{L}=\hat{I}-\hat{L}_{R}^{-1}\hat{Q}_{l}(0)\hat{L}$
with $\hat{Q}_{l}(0)=\hat{N}(0)\hat{N}_{l}^{-1}(0)$ projects on the
null space of the operator $\hat{N}_{l}^{-1}(0)\hat{L}$. Because
the operators $\hat{L}_{R}^{-1}\hat{N}(0)$ and $\hat{L}_{R}^{-1}\hat{G}$
are lower triangular, we can further transform the above equation:
\begin{eqnarray}
 & \left\{ \hat{I}+\lambda_{1}\hat{L}_{R}^{-1}\hat{N}(0)+\hat{L}_{R}^{-1}\hat{N}(0)\hat{N}_{l}^{-1}(0)\hat{G}+\hat{L}_{R}^{-1}\hat{N}(0)\hat{A}(\lambda_{2})\left(\hat{G}+\hat{L}\right)\right\} |V>=\nonumber \\
 & \hat{\Pi}_{L}|V>+\hat{L}_{R}^{-1}\hat{N}(0)\left(\hat{N}_{l}^{-1}(0)+\hat{A}(\lambda_{2})\right)|0>_{info}\label{eq:43}
\end{eqnarray}
and finally: 

\begin{eqnarray}
|V> & +\hat{S}\left(\hat{I}+\lambda_{1}\hat{L}_{R}^{-1}\hat{N}(0)+\hat{L}_{R}^{-1}\hat{Q}_{l}(0)\hat{G}\right)^{-1}\hat{L}_{R}^{-1}\hat{N}(0)\hat{A}(\lambda_{2})\left(\hat{L}+\hat{G}\right)|V>=\nonumber \\
 & \hat{S}\left(\hat{I}+\lambda_{1}\hat{L}_{R}^{-1}\hat{N}(0)+\hat{L}_{R}^{-1}\hat{Q}_{l}(0)\hat{G}\right)^{-1}\hat{\Pi}_{L}|V>\label{eq:44}
\end{eqnarray}
 where symmetry (\ref{eq:37}) was also used. Here again the projector
$\hat{Q}_{l}(0)=\hat{N}(0)\hat{N}_{l}^{-1}(0)$ appears. 

In this approach an important element in the expansion:

\begin{equation}
|V>=\sum_{j=0}^{\infty}\lambda_{2}^{j}|V>^{(j)}\label{eq:45}
\end{equation}
 is the zeroth order approximation $|V>^{(0)}$. This can be calculated
from Eq.\ref{eq:35} in which $\lambda_{2}=0$. For (\ref{eq:28}),

\begin{equation}
\hat{N}(0)=\int d\tilde{x}\hat{\eta}^{\star}(\tilde{x})H(\tilde{x})\label{eq:46}
\end{equation}
 what is a lower triangular operator with respect to projectors (\ref{eq:13}).
It makes that zeroth order problem is possible to solve. Higher orders
approximation are based on the Eq.\ref{eq:44}. Undetermined term,
the vector $\hat{\Pi}_{L}|V>$, can be identified with the zeroth
order approximation:

\begin{equation}
\hat{\Pi}_{L}|V>=\hat{\Pi}_{L}|V>^{(0)}\label{eq:47}
\end{equation}
 This equation can be called the \textit{perturbation principle} according
to which the next approximations to the zeroth order term ($\lambda_{2}=0$)
exclusively depend on the the \textit{perturbation operator.} In Eq.\ref{eq:44}
this is the term that contains the operator $\hat{A}(\lambda_{2})$. 

When Eq.\ref{eq:35} is multiplied by the left-inverse operator $\hat{N}_{l}^{-1}(\lambda_{2})$
- the problem arises, namely - does the obtained Eq.\ref{eq:39} is
equivalent to the initial equation? That the answer is conditionally
positive provides the following reasoning: Multiplying Eq.\ref{eq:39}
by the operator $\hat{N}(\lambda_{2})$ (invertible operation), we
get equation:

\begin{equation}
\left(\hat{Q}_{l}(\lambda_{2})(\hat{L}+\hat{G})+\lambda_{1}\hat{N}(\lambda_{2})\right)|V>=0\label{eq:48}
\end{equation}
where projector $\hat{Q}_{l}(\lambda_{2})=\hat{N}(\lambda_{2})\hat{N}_{l}^{-1}(\lambda_{2})$.
As you can see from the above, Eq.\ref{eq:48} is not equivalent to
Eq.\ref{eq:35}. But Eq.\ref{eq:48} can be equivalently described
as:

\begin{equation}
\left(\hat{I}+\lambda_{1}(\hat{L}+\hat{G})_{R}^{-1}\hat{N}(\lambda_{2})\right)|V>=\hat{\Pi}_{L+G}(\lambda_{2})|V>\label{eq:49}
\end{equation}
 where the projector $\hat{\Pi}_{L+G}(\lambda_{2})=\hat{I}-(\hat{L}+\hat{G})_{R}^{-1}\hat{Q}_{l}(\lambda_{2})(\hat{L}+\hat{G})$.
If, however, 

\begin{equation}
\hat{\Pi}_{L+G}(\lambda_{2})|V>=\hat{P}_{L+G}|V>+(\hat{L}+\hat{G})_{R}^{-1}|0>_{info}\label{eq:50}
\end{equation}
where the projector $\hat{P}_{L+G}=\hat{I}-(\hat{L}+\hat{G})_{R}^{-1}(\hat{L}+\hat{G})$
then Eqs (\ref{eq:48}) and (\ref{eq:49}) are equivalent to Eq.\ref{eq:35}$\clubsuit$. 

To avoid nonphysical solutions we have to use some additional restriction
like (\ref{eq:50}) and to use the perturbation principle like (\ref{eq:47})
resulting from continuity of solutions with respect to the coupling
constant $\lambda_{2}.$ In a sense we have a situation similar to
that which occurs when applying the Galerkin method. The essential
difference is that here any reduction strategy for Galerkin models
is not at least openly used, \cite{Noack (2010)}.

\section{The art of equations and few remarks about one side invertibility
of operators and operator-valued functions}

In the paper presented we used two type of operators describing basic
Eq.\ref{eq:15}: the right and left invertible operators. To the right
invertible operators belongs the operator $\hat{L}$ which usually
describes kinematic properties of the system. An interaction of the
system elements and the exterior world is described by the left invertible
operators: $\hat{N}$ and $\hat{G}$ . These operators here have the
null spaces, responsible for the freedom of the theory, identically
equal to zero. In other words, freedom of a theory is rather exclusively
determined by the operator $\hat{L}$ which is the right invertible
operator. It is seen from the perturbation theory and the considerations
presented here. It is surprising that considered in Sec.5 examples
of the operators $\hat{N}$ are able to approximate polynomial interactions,
which are right invertible operators, but considered as a whole, $\hat{N}$
are left invertible. Moreovere, in addition, it naturally contains
a term which reminds us the operator $\hat{G}$ describing external
or background field. The mass term appears in the next approximation
to the left invertible operator $\hat{N}$. Such connection of the
mass term with the external field recalls the Mach's idea that the
inertial mass of a body is caused by interaction with the rest of
Universe. 

Once again, as we can see from the above, the operators $\hat{N}$
only in polynomial approximations are the right invertible operators.
It seems that left invertibility is a natural property of operators
describing interactive properties of the systems and next to such
properties as symmetry, closeness, see \cite{Hanckow (2010)-1}, will
constitute a significant limitation in the adequate theory search.
On the other hand it can be assumed that the equations like (\ref{eq:15}),
with the only left invertible operators, correspond to the final everything
theory. However the Eq.\ref{eq:15}, in which there are right and
left invertible operators, corresponds to the modern, not-complete
theories needing additional information like boundary conditions to
describe a particular problem. It is also interesting that if we finally
would be able to describe considered, final theory, in a form of inhomogenous
linear equation with a left invertible operator $\hat{A}$ and a given
vector $|\Phi>$:

\begin{equation}
\hat{A}|V>=|\Phi>\label{eq:51}
\end{equation}
the solution, for the vector $|V>$will be unique - in spite of a
structural ambiguity of the operator $\hat{A}_{l}^{-1}$! For this
reason, the left invertibility of the operator $\hat{A}$ is reminiscent
more of both-side invertibility. 

At this point we would also like to notice that expressions given
in Sec.5 like (\ref{eq:28}), (\ref{eq:31}) or (\ref{eq:32}), are
examples of so called (formal) \textit{operator-valued functions}
used in many areas of science. Their simplest examples are polynomial
functions broadly used in quantum field theory. In this paper we have
considered local operator-valued functions like (\ref{eq:28}) and
(\ref{eq:32}). Formulas (\ref{eq:28}), (\ref{eq:31}) suggest their
left invertibility and allows to generate a number of interactions
in a uniform manner. Explicit construction of left-inverse operation
also allowed to obtain compact expressions for successive approximations
with respect to the minor coupling constant $\lambda_{2}.$ The examples
given in Sec.5 lead to equations in which demarcation line among primarily
linear and nonlinear theory was broken, Sec.7, and perhaps, for this
reason, deserve further attention.

In \cite{Hanckow (2010)}, Sec.9, we have used also local operator
valued functions and applied them to closure and regularization purposes. 

The main reason that the almost the same operator-valued functions
$f(\hat{A})$ in \cite{Hanckow (2010)} we treated as right invertible
and here, Sec.5, we treat as left invertible was the undefined status
of the functions like (\ref{eq:28}) or rather operator-valued function
$\frac{\hat{I}}{\hat{I}-\lambda_{2}\hat{\eta}(\tilde{x})}$. Happy
coincidence is that, under some additional conditions as the (\ref{eq:48}),
this leads to similar results. 

As additional literature we recommend: \cite{Przew (1988)}, \cite{Voicu (1991)}
and \cite{Efimov (1977)}.

\section{When usual derivatives and integrals are not necessary}

Since the advent of a Newton and Leibniz era derivatives and integrals
are an essential tool for the description of nature. With their help,
almost every equation of physics and technology is recorded. So we
proceed when to their description we use the generating functionals
or vectors, see \cite{Rzew (1969),Efimov (1977)}. But tradition is
not always a good guide, see the New Testament. This is what causes
that in case of the generationg functionals and vectors we must not,
and I believe that we should not use derivatives, is their \textbf{formal
character}. It makes no sense to talk about changes of the formal
quantities. In the papers, instead derivatives and integrals we have
used the operators $\hat{\eta}(\tilde{x})$ and $\hat{\eta}^{\star}(\tilde{x})$
that resemble derivatives only in the fact that they are one-side
invertible operators. No Leibniz identity analoge. The canonical derivatives
and integrals are used on the ``micro'' level of description in
the paper represented by Eqs (\ref{eq:16}). To imagine the complications
which can occur when on the level of \textquotedbl{}macro\textquotedbl{}
we use traditional derivatives, or rather their functional analogs,
we present them by means of operators $\hat{\eta}(\tilde{x})$ and
$\hat{\eta}^{\star}(\tilde{x})$. We have:

\begin{equation}
\delta/\delta\eta(\tilde{x})\Leftrightarrow\sum_{n=1}^{\infty}\sum_{k=0}^{n}\hat{\eta}^{\star k}\hat{\eta}(\tilde{x})\hat{\eta}^{k}\hat{P}_{n}\Rightarrow\hat{\eta}(\tilde{x})\label{eq:52}
\end{equation}
In the post-Newton-Leibniz era, in certain areas of science, much
simpler variables than the canonical variables - should be used. On
the other hand, used type of combination of variables reminds us a
classical-quantum description of the phenomena in which, at the same
time, derivatives and action integrals appear. 

The above modification of the usual calculus (Newton and Leibniz (including
fractional derivatives and so on)) reminds us of the transition from
commutative space-time to noncommutative one. But this is only a formal
similarity, because here we get rid of the generating vectors - the
derivatives in the Newton and Leibniz sense, and there points and
moments are got rid of the space, \cite{Heller (2008)}. In other
words, in our approach, at the micro-level, Newton's and Leibniz's
calcules is used, and a departure from it is done at the macro-level.
We believe that the micro-level is rather closer to point-spaces what
it is hidden under the term: fine-grained structure. The coarse-grained
structure are related to some smoothing or averaging procedures. And
here comes the brand new, paradoxical phenomenon: to describe the
average quantities, you must also consider correlations between them.
In other words, we consider infinite collection of n-pi. If this collection
of n-pi is considered in the free Fock space, the description of equations
for n-pi is more effective and even can be used to describe quantum
phenomena, see, e.g.,\cite{Hanckow (2011),Hanckow (2010)}. This last
case is probably a consequence of two facts: quantum phenomena belong
to micro-level but measurments on them belong to macro-level and the
free Fock space allows us to connect these two levels together. It
is hoped that the same space will allow you to connect QM and GTR
(sic).

\end{document}